\documentclass[aps,pra,twocolumn,superscriptaddress]{revtex4}
\usepackage{amsmath,epsfig,mathrsfs,graphicx,amssymb,slashed}

\begin{document}
\title{Analysis of assumptions of recent tests of local realism}
\author{Adam Bednorz}
\email{Adam.Bednorz@fuw.edu.pl}
\affiliation{Faculty of Physics, University of Warsaw, ul. Pasteura 5, PL02-093 Warsaw, Poland}
%\date{\today}

\begin{abstract}
Local realism in recent experiments is excluded on condition of freedom or randomness of choice combined with no signaling between observers by implementations of simple quantum models. Both no-signaling and the underlying quantum model can be directly checked by analysis of experimental data.  For particular tests performed on the data, it is shown that two of these experiments give the probability of the data under no-signaling (or choice independence in one of them) hypothesis at the level of 5\%, accounting for the look-elsewhere-effect, moderately suggesting that no-signaling is violated with 95\% confidence. 
On the other hand the data from the two other experiments violate the assumption of the simple quantum model. Further experiments are necessary to clarify these issues and freedom and randomness of choice.
\end{abstract}

\maketitle 

\section{Introduction}

The question of local realism of quantum observations has been raised by Einstein, Podolsky and Rosen \cite{epr}.
Much later Bell turned this question into verifiable inequality satisfied by local realism and showed a theoretical quantum counterexample.
The Bell theorem states that quantum mechanics violates local realism in the form of testable inequalities \cite{bell,chsh,eber}. Two observers are separated and make their measurements mutually spacelike or at least compatible. In the test it is essential that the observers make a free random choice of what to measure and complete the readout spacelike.  Only then the violation of a Bell-type inequality \cite{chsh,eber} refutes local realism because, comparing to local real models,  one assumes that the readout of one observers does not depend on the other one. In addition, experimentally feasible quantum examples violating local realism are based on simple mathematical models, also assumed.

For decades, Bell tests were realized usually with photons \cite{belx1a,belx1b,belx1c,belx1g,belx2a,belx2b}.
However, the advantage of the Bell-type tests is that
no specific system is required, it only has to fit the simple quantum few-state approximation, which resulted in setups across nearly 
all branches of physics \cite{belx1d,belx1e,belx1f,belx1h}, see the review \cite{genov}. Unfortunately, among various problems the most significant appeared lack of sufficient  distance (locality loophole), imperfect detection (detection loophole) \cite{loop1,loop2,graft} e.g. a fraction of particles are lost and predetermined (often 
fixed) choices (random or free choice hypothesis). Loopholes allow for a local realistic model \cite{loopmod1,loopmod2}. Bell test is also stronger than entanglement criteria \cite{ent} because the latter rely on 
specific representation of observables in quantum space while the former only refers to outcomes of measurements. It is also stronger 
than steering where one assumes quantum representation of observables at one of the parties \cite{eprst1,eprst2,eprst3}.
However, Bell examples do not violate a weaker context-related local realism \cite{gran}.

In the recent Bell tests performed in Delft \cite{hensen}, NIST \cite{nist}, Vienna \cite{vien} and Munich \cite{munch}, claimed as loophole-free,  violation of local realism is claimed with high confidence level (assuming local realism, the probability of the data is 4\% in \cite{hensen}, $\sim 10^{-7}$ in \cite{nist}, $\sim 10^{-31}$ in \cite{vien} and $\sim 10^{-9}$ in \cite{munch}). At the same time, the data from these experiments can be used to verify the above mentioned assumptions, which is the purpose of the hereby analysis.

No-signaling or microscopic causality means that no information about a free choice can be transmitted to the other observer. The Bell theorem alone does not specify when no-signaling is valid, just assumes it. This is why no-signaling is usually adopted from relativity. An observation cannot be affected by a disturbance located at a point in spacetime such that  the distance by the time difference is greater than speed of light (spacelike). Of course no-signaling is valid not only in quantum mechanics but any theory consistent with special relativity, although usually no-signaling is rather postulated \cite{wight} than proved or derived.
Interestingly, some of Wightman axioms can be proved, e.g. relativistic invariance of the vacuum \cite{b13}. In quantum theory it means that spacelike separated freely chosen disturbance and measurement cannot affect each other, they are compatible (commuting).
Even without Bell theorem, freedom of choice is in conflict with objective realism in relativistic quantum field theory, which puts special relativity at stake \cite{b16}.
The absence of relativistic signaling is only warranted by the setup in space and time such that it would have to be faster than light, forbidden by special relativity. Therefore confirming signaling in such experiments would also mean violation of relativity or delay in the factual readouts. Moreover, the experiments rely on a simple quantum model of an entangled state of two qubits (two-state systems, either polarizations of photons or fine-tuned solid state impurities) with the measurements corresponding to pseudospin readout along a freely chosen axis.
Randomness of choice is at present warranted by electronics,  based on random number generators, supported by pseudorandom input, so one has to trust devices. The trust could be less important if the time for the choice is longer (of the order of seconds) but it would require e.g. Earth-Moon distance (trusting relativity), beyond current technology \cite{wise}.

Although simple quantum models predict preserving no-signaling with free choice while violating of local realism, all these properties are subject to experimental tests. 
In the paper, each of the experiments is examined separately, Delft, NIST, Vienna and Munich, accordingly. They are very different, e.g. the distance between observers was $1.3$km in Delft, $180$m in NIST,  $58$m in Vienna and $398$m in Munich, the recored data have different volume and processing protocols. It is found by specific tests that assuming no-signaling the probability of the data is only about 5\% in Delft and NIST, taking into account look-elsewhere effect. The Delft data are also currently examined in other reports \cite{khlar}. The data from Vienna and Munich show 
incompatibility with the simple quantum model. For all experiments the time left for the choice is trusted.  We summarize all findings in the Discussion.

\section{Local realism}

\begin{figure}[h]
%\vspace*{2cm}
%\hspace*{-4cm}
\includegraphics[scale=.3]{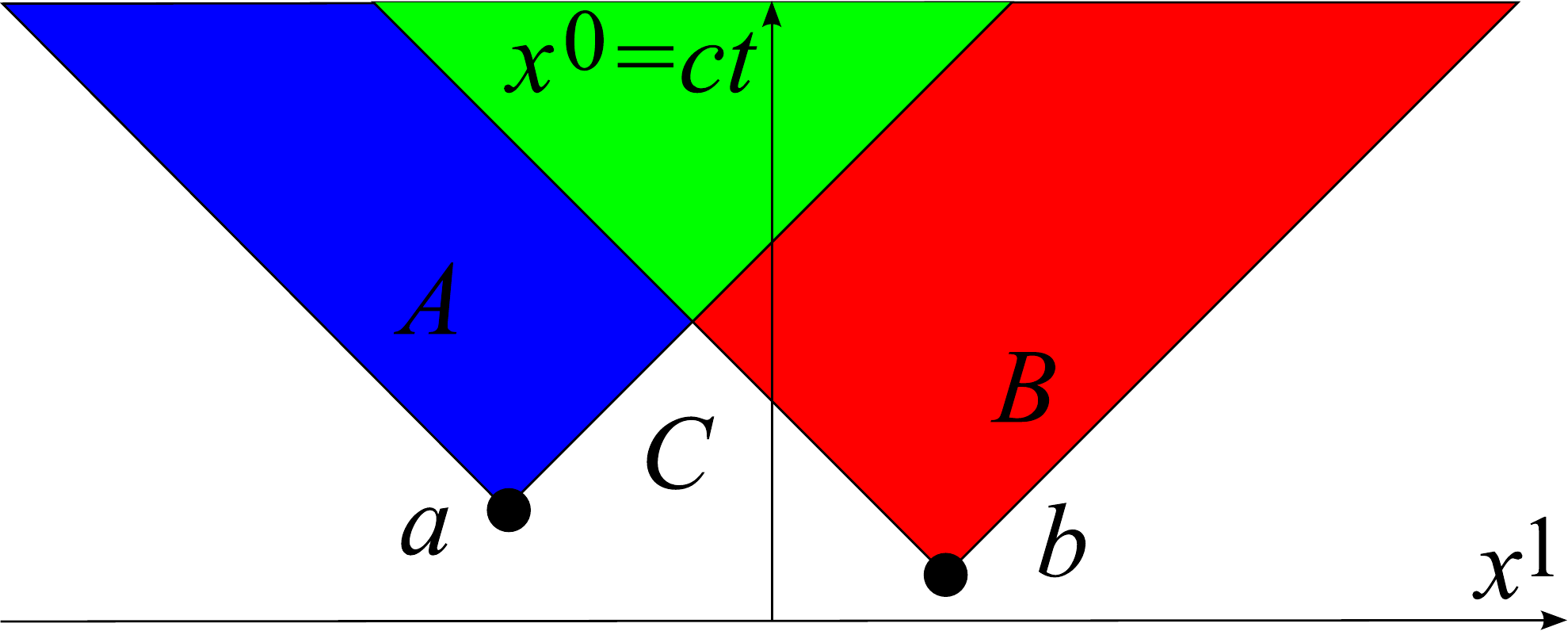}
\caption{No-signaling in the case of two choices $a$ and $b$ and three measurements $A$ (blue), $B$ (red) and $C$ (white) lying spacelike with respect to $b$, $a$ and both, respectively, bounded by the speed of light $|\Delta x^1|<\Delta ct=\Delta x^0$ [$\Delta x= x(\mathrm{readout})-x(\mathrm{choice})$]. According to the principle the outcome of $A$ cannot depend on $b$, $B$ on $b$ and $C$ on  both. A measurement in green region can depend on both $a$ and $b$.}
\label{sii}
\end{figure}

Let us start recalling predictions of local realism based on relativistic no-signaling. Randomly chosen influences (e.g. qubit rotations) $a_x$ ($=0,1$ in a dichotomic choice) and measurements of $A_x$ are located in spacetime points $x=(x^0=ct,\vec{x})$, with time $t$, speed of light $c$ and spatial position $\vec{x}$.
According to no-signaling principle, the probability $p(A_x|a_y)$ of the measurement of $A$ at $x$ while the choice $a$ at $y$ has been made
cannot depend on $a$ if $x-y=z$ is spacelike, namely $|z^0|<|\vec{z}|$, which visually means being outside of the causality cone, $|z^0|\geq |\vec{z}|$ and $z_0>0$. More generally $p(A_x,B_y,C_z,...|a_u,b_v,...)$ will not depend on $a$ if $x-u$, $y-u$, $z-u$ are spacelike, and similarly with $b$. Local realism in the situation depicted in Fig. \ref{sii} means that
$x-v$, $y-u$ and $z-u$, $z-v$ are spacelike (optionally $u^0,v^0>z^0$) and there exits a joint probability distribution for all choice-dependent outcomes
$p(\{A_a\},\{B_b\},C)$ with $A_a$ meaning outcome $A$ at the choice $a$. No-signaling means that $A$ is independent of $b$. Such a joint probability can never be determined exactly (because we cannot measure correlations between different choices) but its positivity implies bound on measurable correlations \cite{bell,chsh,eber}. We will drop spatiotemporal indices when there is no ambiguity.

\section{Delft experiment}

In the Delft experiment \cite{hensen} we have the situation depicted in Fig. \ref{sii}.
The spatial-temporal distances $x-v$ and $y-u$ were 1280m/4.27$\mu$s, $z-u$ were 493m/1$\mu$s, and $z-v$ was 818m/1$\mu$s and they are all spacelike. The experiment involves entangling electron spin in a nitrogen-vacancy center with a photon and swapping the entanglement with another spin-photon pair by detecting the suitably superposed pair of photons.
The outcome $C$ corresponds to the specially filtered pair of clicks due to incoming photons while $A$ and $B$ correspond to outcome of spin measurement in the basis chosen by $a$ or $b$. The choice is performed by microwave pulse rotating the spin.

Here $C=1$ heralds successful entanglement swapping ($0$ otherwise). The choices $a$ and $b$ are independent and dichotomic, assigned $0$ or $1$. The measured values of spin states, $A$ and $B$ are $+1$ or $-1$. The value of $C$ cannot depend on either $a$ or $b$ and so the probability $p(C=1)$ should be the same for all values of $a$ and $b$. Taking the data from the paper \cite{hensen}, Fig. 4(a), one can find this probability times the total number of attempts for every combination of $a$ and $b$, see Table \ref{tab1} left.

\begin{table}
\begin{tabular}{c|c|c|c}
$C=1$&$0$&$1$&$a$\\
\hline
$0$&$53$&$62$&\\
\cline{1-3}
$1$&$79$&$51$&\\
\cline{1-3}
$b$&\multicolumn{3}{c}{}\\

\end{tabular}
\begin{tabular}{c|c|c|c}
$C=1$&$0$&$1$&$a$\\
\hline
$0$&$175$&$195$&\\
\cline{1-3}
$1$&$218$&$159$&\\
\cline{1-3}
$b$&\multicolumn{3}{c}{}\\
\end{tabular}
\caption{Heralding events $Cab=p(C=1|a,b)N$  for all combinations of $a$ and $b$ times the number of attempts $N$ in (left) and without (right) the heralding window.}\label{tab1}
\end{table}

It is clear that the number of heralding events for choices $ab=01$, namely $Np(C=1|a=0,b=1)=C01=79$ ($N$ is the total number of heralding events) is much different from $Np(C=1|a=1,b=1)=C11=51$. These probabilities should be equal if the random number generators work properly. 
Of course the differences can be due to statistical error, represented by $P$-value.
Equal probability (no signaling) is our null hypothesis which, similarly to local realism will be assigned a $P$-value -- the probability that it is true for the experimental  statistics (counting probability of actual or larger deviation from predicted average).
Given a test statistic, the $P$-value is defined as the probability, under the assumption of the hypothesis of obtaining a test statistic value equal to or more extreme than what was actually observed.

 Summing $01$ and $11$ events we obtain $130$ which gives $P$-value $0.0175$ (twice the probability of less of equal successes than $51$ out of $130$ for binomial distribution and $1/2$ success probability). However, by the look-elsewhere effect in terms of Bonferroni correction \cite{bonf} -- standard, accepted and well-understood statistical
methodology, we could take any column or row (for $a$ or $b$ fixed) so in fact it is 4 times larger, $0.07$ giving $93\%$ confidence level for violation of no-signaling principle. The look-elsewhere effect is common in particle physics, where the data must be normalized for the number of places searched in which a fluctuation could be observed \cite{lee}. Look-elsewhere effect reflects the fact that large fluctuation is possible in a sufficiently large set of data. Thanks to accounting for this effect, particle physics avoided premature announcement of false discoveries.
 Moreover, in the reported data \cite{hensen} the coincidence $C=1$ is filtered by selecting certain time windows for detected photons. Taking full data \cite{hensenraw} and relaxing the heralding window time constraints to get a result in Table \ref{tab1}(right).

Then $79\to 218$ and $51\to 159$ (the distribution is roughly the same), giving the $P$-value $0.00276$, or, with the factor $4$, $0.011$  so the confidence level is around $98.9\%$, higher than Bell violation ($P$-value $0.039$).
Moreover, the independence of $a$ and $b$ is violated. Namely, if $0$ is chosen with a probability $p_a$ and $1$ with $q_a$, $p_a+q_a=1$ and similarly for $b$ then still independence requires that 
\begin{equation}
C00\:C11=Np_ap_bq_aq_b=C01\:C10=Np_aq_bq_ap_b
\end{equation} 
while $53\cdot 51$ differs much from $79\cdot 62$. Performing Pearson's $\chi^2$ test of independence one obtains $P$-value 
$0.0077$ while relaxing the heralding window one gets $0.0013$. Therefore the data suggest either moderate signaling or statistical dependence.

\begin{table}
\begin{tabular}{c|c|c|c}
$C=1$&$0$&$1$&$a$\\
\hline
$0$&$420$&$386$&\\
\cline{1-3}
$1$&$361$&$373$&\\
\cline{1-3}
$b$&\multicolumn{3}{c}{}\\

\end{tabular}
\begin{tabular}{c|c|c|c}
$C=1$&$0$&$1$&$a$\\
\hline
$0$&$193$&$179$&\\
\cline{1-3}
$1$&$176$&$182$&\\
\cline{1-3}
$b$&\multicolumn{3}{c}{}\\
\end{tabular}
\caption{Heralding events $Cab=p(C=1|a,b)N$  for all combinations of $a$ and $b$ times the number of attempts $N$ for $\psi_+$ (left) and $\psi_-$ (right) without the heralding window in the second run.}\label{tab2}
\end{table}

The data from the second run \cite{hensen2} do not show significant signaling (or random number dependence) but a moderate evidence is present in the raw data \cite{hensenraw2} with the coincidence window relaxed. The run takes coincidences from different photodetectors ($\psi_-$ state) and the same one ($\psi_+$), see the Table \ref{tab2}. Now the total number of $00$ and $01$ events are $613$ and $537$ giving a $P$-value of $0.027$
but again accounting for all rows and columns, it must be multiplied by $4$ giving a $P$ of $0.11$. This is larger than the Bell test $P$-value reported in \cite{hensen2} ($0.061$) but the differences in the distribution between $\psi_+$ and $\psi_-$ are similar. Moreover, the total number of events is now
$2270$ (with the majority of $\psi_+$, not considered in the first run) which is   $3.04$ times $747$ from the first run. Assuming the look-elsewhere effect (Bonferroni correction \cite{bonf}) the value from the first run must be multiplied by $4.04$ giving $0.045$, still of the same order to the Bell $P$-values for local realism of both runs.

Summarizing the Delft experiment, although the statistics are small, it shows moderate violation of either no-signaling or random number independence, visible rather in the first run than in the second.

\section{NIST experiment}

The NIST experiment \cite{nist} differs considerably from Delft because it does not involve the middle observation point $C$, see Fig. \ref{sii}
while the outcomes at observation points A and B are $1$ (denoted $+$ in the paper) and $0$ instead of $+1$ and $-1$. The rest of the setup is analogous. The distance $x-v$ and $y-u$ is $184.9$m while the time difference is $\sim 950$ns. It depends on the pulse number.
In the experiment one outcome is measured in 800 subsequent and synchronized time intervals ($12.6$ns), corresponding to laser pulses, while
the local realism is tested for a selected 16 of them (counted from 28 for A and 37 for B). 
Only part of them are spacelike (already the last of selected pulses are not spacelike) and NIST authors take into account only spacelike pulses in their analysis of local realism. However, in the discussion below we analyze all pulses, to check if there is at all signaling (even subluminal). The experiment generates entangled pairs of photons. In the test $A=1$ or $0$ (similarly $B$) depending of where a photon was registered in the appointed photodetector (clicked) in preselected time window. The choice is made by Pockels cells which rotate the polarizations of incoming photons.

Assuming fair random number generators, $p(A=1|ab)$ cannot depend on the choice $b$ while $p(B=1|ab)$ on the choice $a$. Then e.g. $p(A=1|a,b=0)=p(A=1|a,b=1)$ which is the null hypothesis to be checked. 
In the supplemental material of the paper \cite{nist} this has been presented in Table S-II, for the run which showed the lowest $P$-value for the hypothesis of local realism.  For the data in Table S-II, the $P$-value for the hypothesis of no-signaling is $\sim 1$, so it does not provide evidence against no-signaling. In the Table S-III the hypothesis is tested for all runs, showing $P$-values distributed randomly. Even very small $P$-value is then only a look-elsewhere effect, fair interpretation requires multiplying $P$ by the number of checks or runs (120 in that case). However, it is questionable to mix all runs because only for those where local realism has been violated with very low $P$-value
testing signaling is critical. One can of course check also other runs but lack of signaling can be there correlated with larger $P$-value for local realism. For this reason we focus mainly on the last run, denoted in \cite{nist} as Classical XOR3, where local realism has been violated with $P$-value $\sim 10^{-7}$, commenting also other runs later.

Still, the analysis in the paper \cite{nist} has been done for appropriately filtered data. Namely, the observation time is converted to the pulse number $0-800$ and phase ($0-160$). This is done by dividing the period of the synchronizing laser into 800 equal pulses so photon detection time is tagged by pulse number and remaining phase. The 16 pulse numbers relevant for the experiment start at $28$ and $37$ for A and B respectively, because this is the first pulse affected by the choice while the phase window is $90\pm 4$ for A and $125\pm 5$ for B. Since the raw data are available \cite{nistraw}, one can extend the analysis removing the pulse and phase window. One should be aware of three important factors. First, pulse number $\sim 800$ 
are certainly already inside the reach of speed of light. In an ideal Bell test the setup should not allow signaling at any speed (conventional communication would be much slower due to material properties of fiber cables) but in practice some subluminal signaling is possible for pulses with large numbers. Second, detection events, $A,B=1$ are most likely accumulated in the predefined time window so one has to consider separately events close to the peak and far away. Third, the look-elsewhere-effect will be taken into account when the $P$-value for the (post-)selected window will be scaled to full window. In this way one avoids artificially small $P$-values for particular values of pulse and phase.

For each window selection, we have 8 pairs of event numbers to compare, either $A=1$ or $B=1$ and all settings $ab$.
Below we present such a comparison for pulses $28-800$ and $37-800$ and the phase window inside (Table \ref{tabn1}) or outside (Table \ref{tabn2}) $90\pm 16$ and
$125\pm 20$ for A and B, respectively (4 times the original window).  The starting pulse corresponds to the first one where the choice 
is affecting the setup. The window is chosen to separate the peak events from the uniform background. As the border is somewhat arbitrary, we will make a later correction to take into account possible different windows to prevent accidentally low $P$-value from special choice of window boundary.

\begin{table}
\begin{tabular}{c|c|c|c}
$A=1$&$0$&$1$&$a$\\
\hline
$0$&$502339$&$503113$&\\
\cline{1-3}
$1$&$505163$&$502146$&\\
\cline{1-3}
$b$&\multicolumn{3}{c}{}\\

\end{tabular}
\begin{tabular}{c|c|c|c}
$B=1$&$0$&$1$&$a$\\
\hline
$0$&$132496$&$132503$&\\
\cline{1-3}
$1$&$132702$&$132816$&\\
\cline{1-3}
$b$&\multicolumn{3}{c}{}\\
\end{tabular}

\caption{Number of detection events $(A,B=1)$ for all combinations of $a$ and $b$ times \emph{outside} the assumed phase window, see text.}\label{tabn1}
\end{table}

\begin{table}

\begin{tabular}{c|c|c|c}
$A=1$&$0$&$1$&$a$\\
\hline
$0$&$1629300$&$1683954$&\\
\cline{1-3}
$1$&$1629753$&$1683455$&\\
\cline{1-3}
$b$&\multicolumn{3}{c}{}\\

\end{tabular}
\begin{tabular}{c|c|c|c}
$B=1$&$0$&$1$&$a$\\
\hline
$0$&$1534134$&$1531788$&\\
\cline{1-3}
$1$&$1598892$&$1601048$&\\
\cline{1-3}
$b$&\multicolumn{3}{c}{}\\
\end{tabular}
\caption{Number of detection events $(A,B=1)$ for all combinations of $a$ and $b$ times \emph{inside} the assumed phase window, see text.}\label{tabn2}
\end{table}

Outside of the presumed window we have choice-conditioned number of events $N(A=1|00)=502339$ and $N(A=1|01)=505163$ with the difference $2824$ against the sum $1007502$.
Assuming Gaussian distribution we get raw $P$-value of $0.0049$. However due to look-elsewhere-effect such a difference could appear for any pair of choices so we have to multiply it by $4$ and rescale by the cut out window factor $160/(160-32)$ giving $0.0245$
(assuming flat distribution the outside detections should occur also inside but they are covered by the peak). Additionally it could occur \emph{inside} the window (with many more events) so Bonferroni correction \cite{bonf} gives another factor 2 (look-elsewhere for two applied tests) and finally $P=0.049$.

\begin{figure}[h]
\includegraphics[scale=.4]{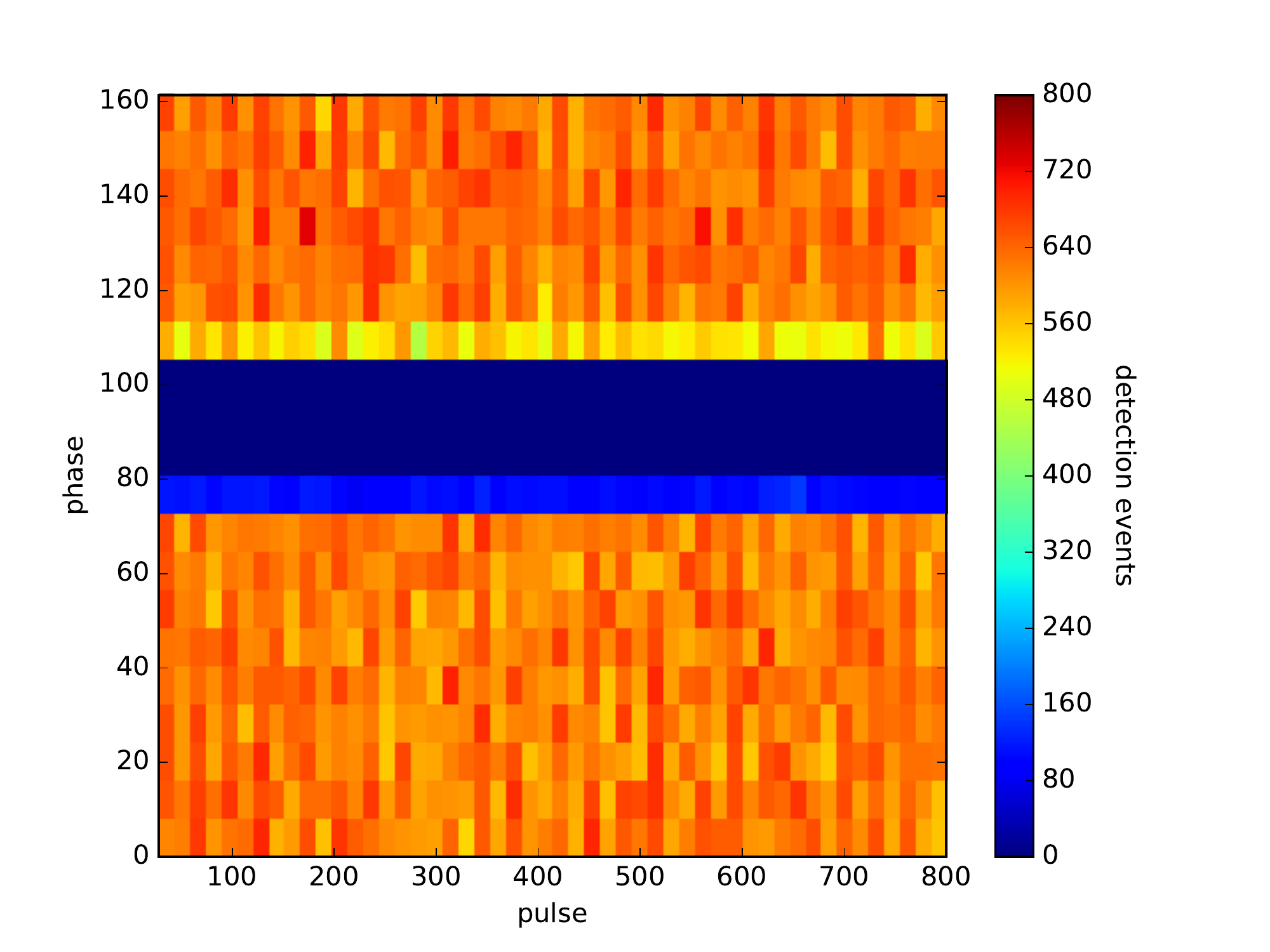}
\includegraphics[scale=.4]{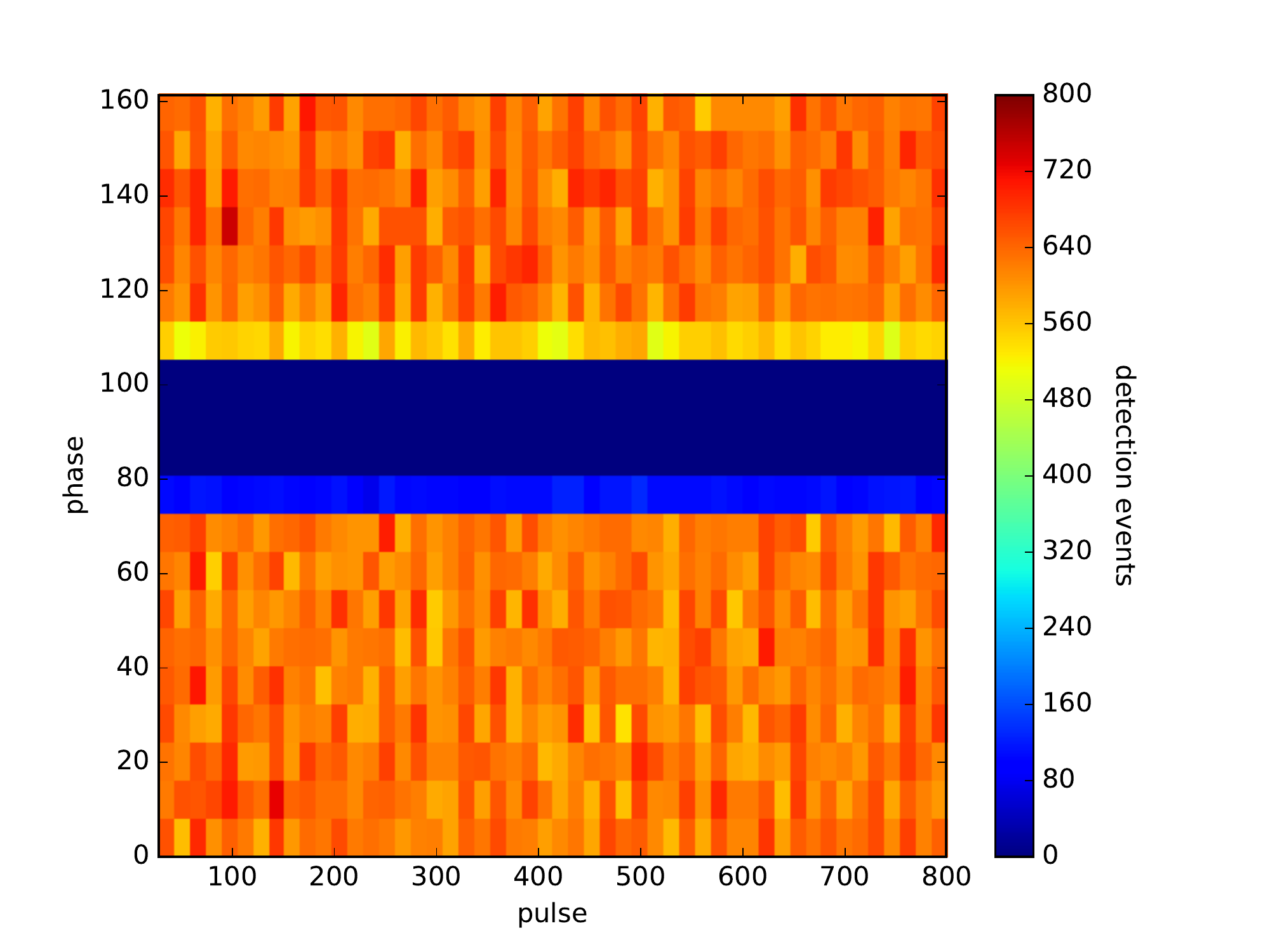}
\caption{The distribution of detection events outside of the $90\pm 16$ phase window for $ab=00$ (top) and $01$ (bottom) grouped in pulse $\times$ phase bins $50\times 20$.}\label{hist2d}
\end{figure}

We have also checked all the range of pulses and phases and no large deviations appear. For the particular case, $ab=00,01$, 
the distribution of detection events seem uniform, independent of a particular pulse of phase, constituting a usual dark count background, see Fig. \ref{hist2d}.  At present, $P\simeq 5\%$ does not necessarily mean that the signaling, even if confirmed, is superluminal because most of the pulses lie within the reach of speed of light. What is also strange, the deviation appears long after the choice was switched off ($28+16$ pulses). The choice means that the photon polarization is changed by Pockels cells only during 16 pulses and the cells are otherwise off. However, within the present data it is impossible to point out a specific time or pulse when the signaling takes place -- it would require much more data.

As regards other runs, we have analyzed also Classical XOR 1 and 2, performed shortly before XOR3 but with larger $P$-values for local realism in range $0.001-0.01$. From the analysis, we have checked if the low $P$-value obtained in XOR3 is also present in XOR1 and XOR2 data, also taking the same outside window. The results are shown in Tables \ref{tab-xor2} and \ref{tab-xor1}. On can directly check that also $P$-values for no-signaling are larger and the most significant cases are different than XOR3.
In particular, for XOR1, one has $N(B=1|00)=132471$ and $N(B=1|10)=131117$, giving raw $P=0.00836$, multiplied by $8$ (2 cases for A and B and 2 runs) $P=0.0688$. No other factors are necessary, because the test has been set up in advance, to check consistency with XOR3 results. Larger $P$-value and different case may be caused by (i) statistical error and (ii) sensitivity of the setup to the alignment changing the $P$-values for local realism. Note that the total event numbers are significantly different than in XOR3, which reminds that we cannot simply mix up the results from different runs.

\begin{table}
\begin{tabular}{c|c|c|c}
$A=1$&$0$&$1$&$a$\\
\hline
$0$&$496775$&$497351$&\\
\cline{1-3}
$1$&$496370$&$497454$&\\
\cline{1-3}
$b$&\multicolumn{3}{c}{}\\

\end{tabular}
\begin{tabular}{c|c|c|c}
$B=1$&$0$&$1$&$a$\\
\hline
$0$&$131581$&$131834$&\\
\cline{1-3}
$1$&$131198$&$131041$&\\
\cline{1-3}
$b$&\multicolumn{3}{c}{}\\
\end{tabular}

\caption{Number of detection events $(A,B=1)$ for all combinations of $a$ and $b$ times outside the assumed phase window, as in Table \ref{tabn1}, for Classical XOR2}\label{tab-xor2}
\end{table}

\begin{table}
\begin{tabular}{c|c|c|c}
$A=1$&$0$&$1$&$a$\\
\hline
$0$&$499281$&$499439$&\\
\cline{1-3}
$1$&$499640$&$500829$&\\
\cline{1-3}
$b$&\multicolumn{3}{c}{}\\

\end{tabular}
\begin{tabular}{c|c|c|c}
$B=1$&$0$&$1$&$a$\\
\hline
$0$&$132471$&$131117$&\\
\cline{1-3}
$1$&$132430$&$131507$&\\
\cline{1-3}
$b$&\multicolumn{3}{c}{}\\
\end{tabular}

\caption{Number of detection events $(A,B=1)$ for all combinations of $a$ and $b$ times outside the assumed phase window, as in Table \ref{tabn1}, for Classical XOR1}\label{tab-xor1}
\end{table}

Summarizing the NIST experiment, it shows moderate violation of no-signaling but the possible signaling may be also subluminal. However, no clear boundary for signaling is visible in the data.

\section{Vienna experiment}

The experiment in Vienna is analogous to NIST (entangled photons, click photodetection and choice by Pockels cells), also corresponding to Fig. \ref{sii} without $C$. Here the spatial/temporal distances $x-v$ and $y-v$ are 58m/180ns (spacelike).

\begin{figure}
\includegraphics[scale=.5]{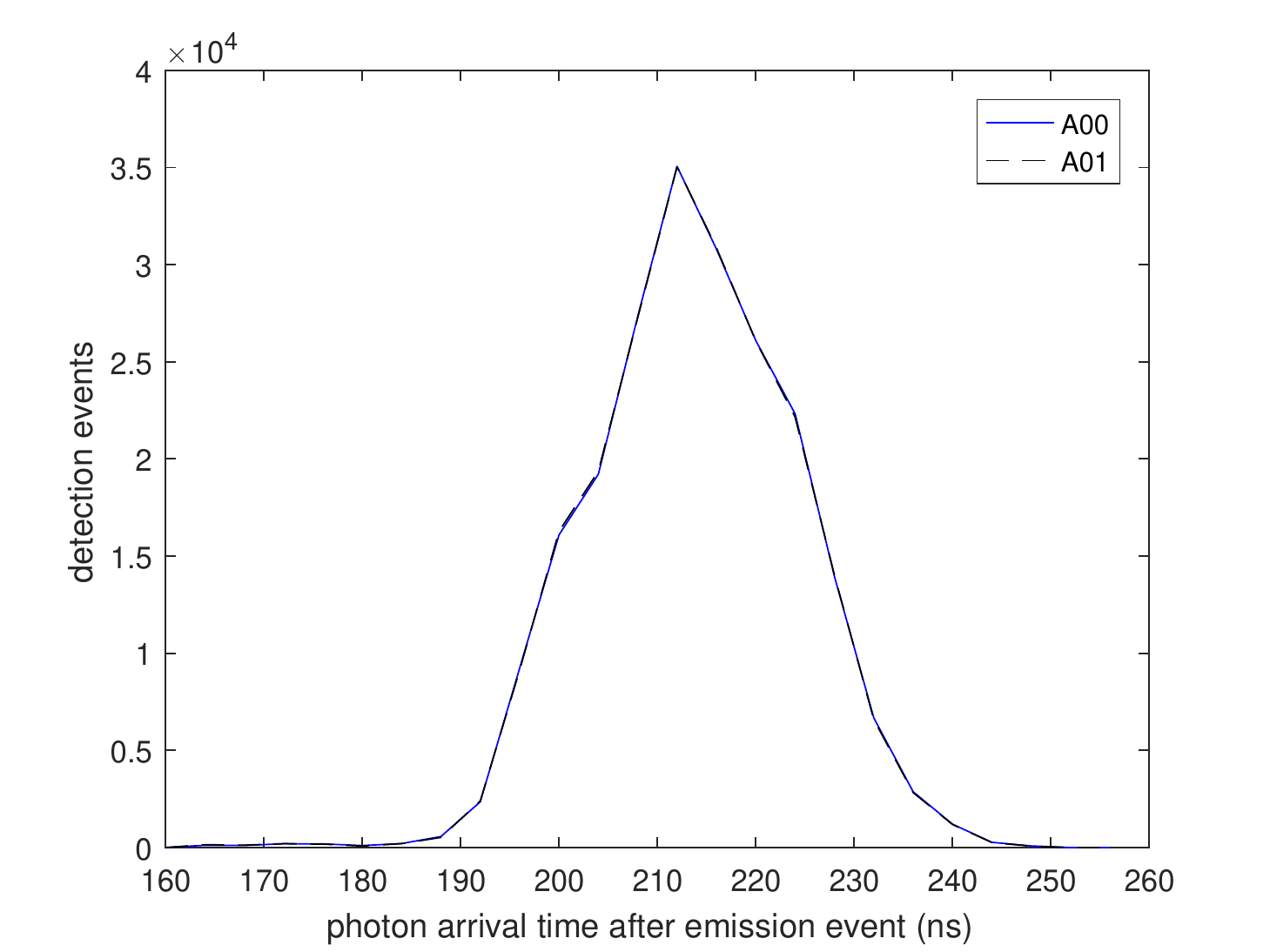}
\includegraphics[scale=.5]{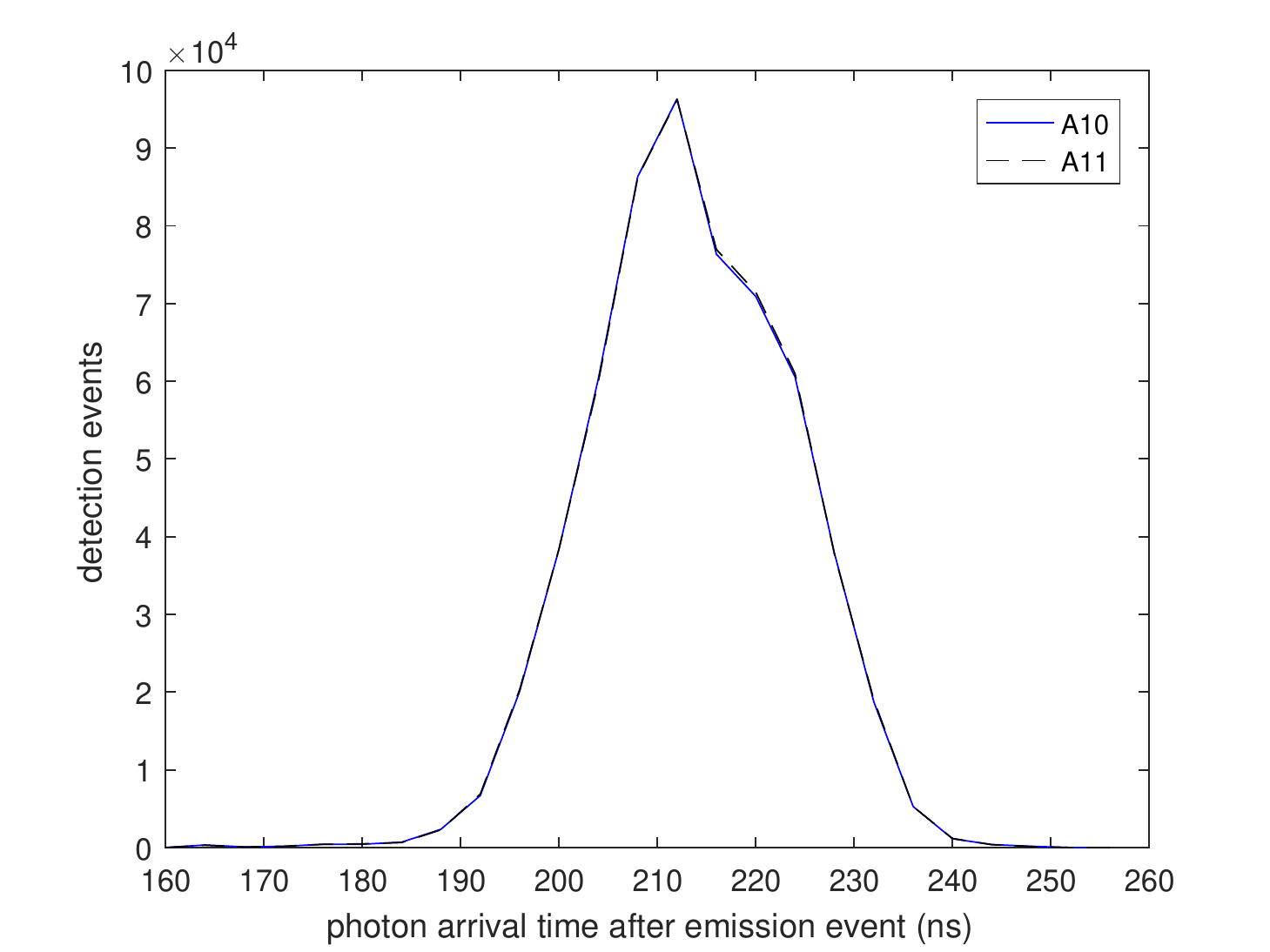}
\includegraphics[scale=.5]{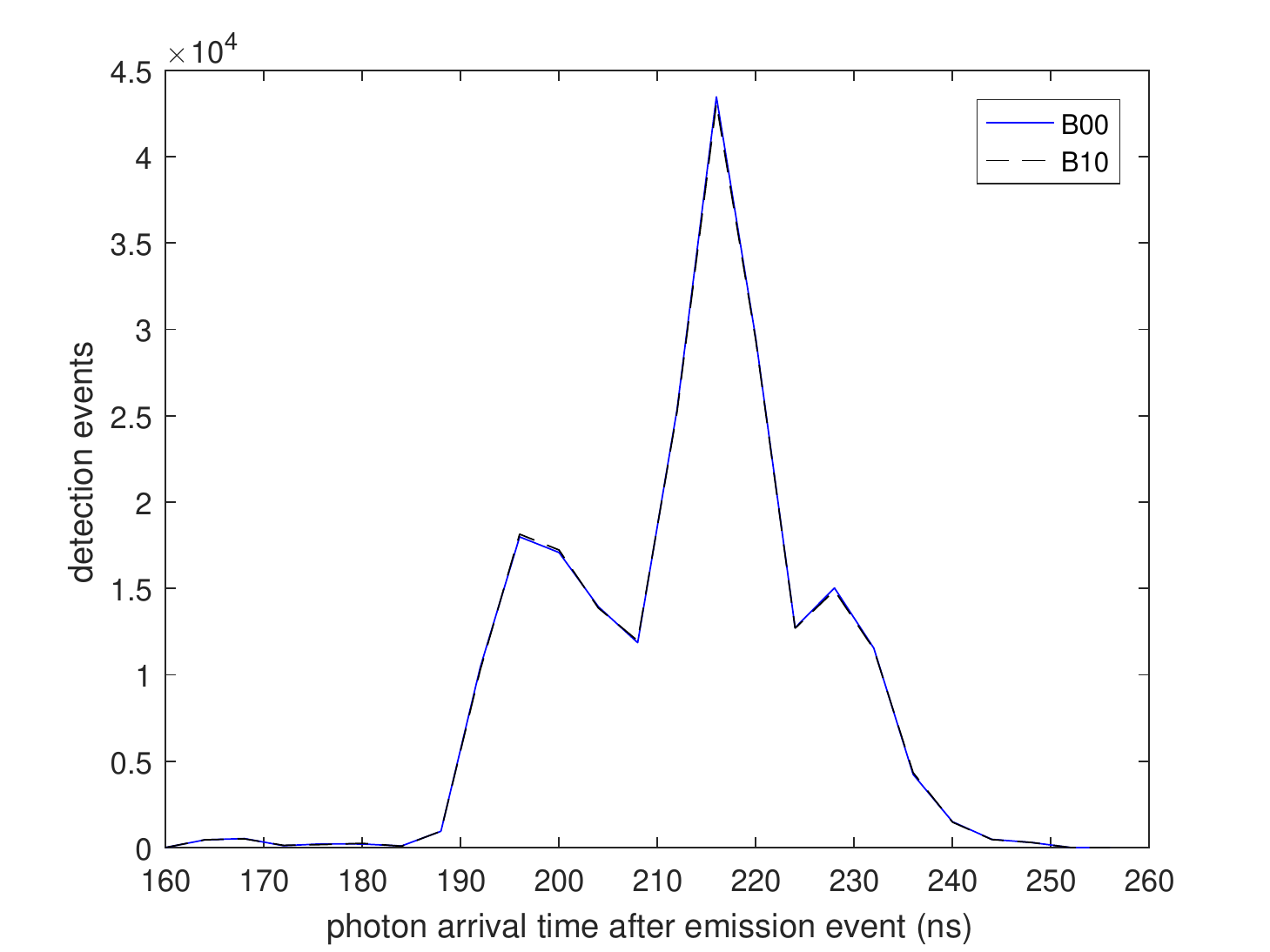}
\includegraphics[scale=.5]{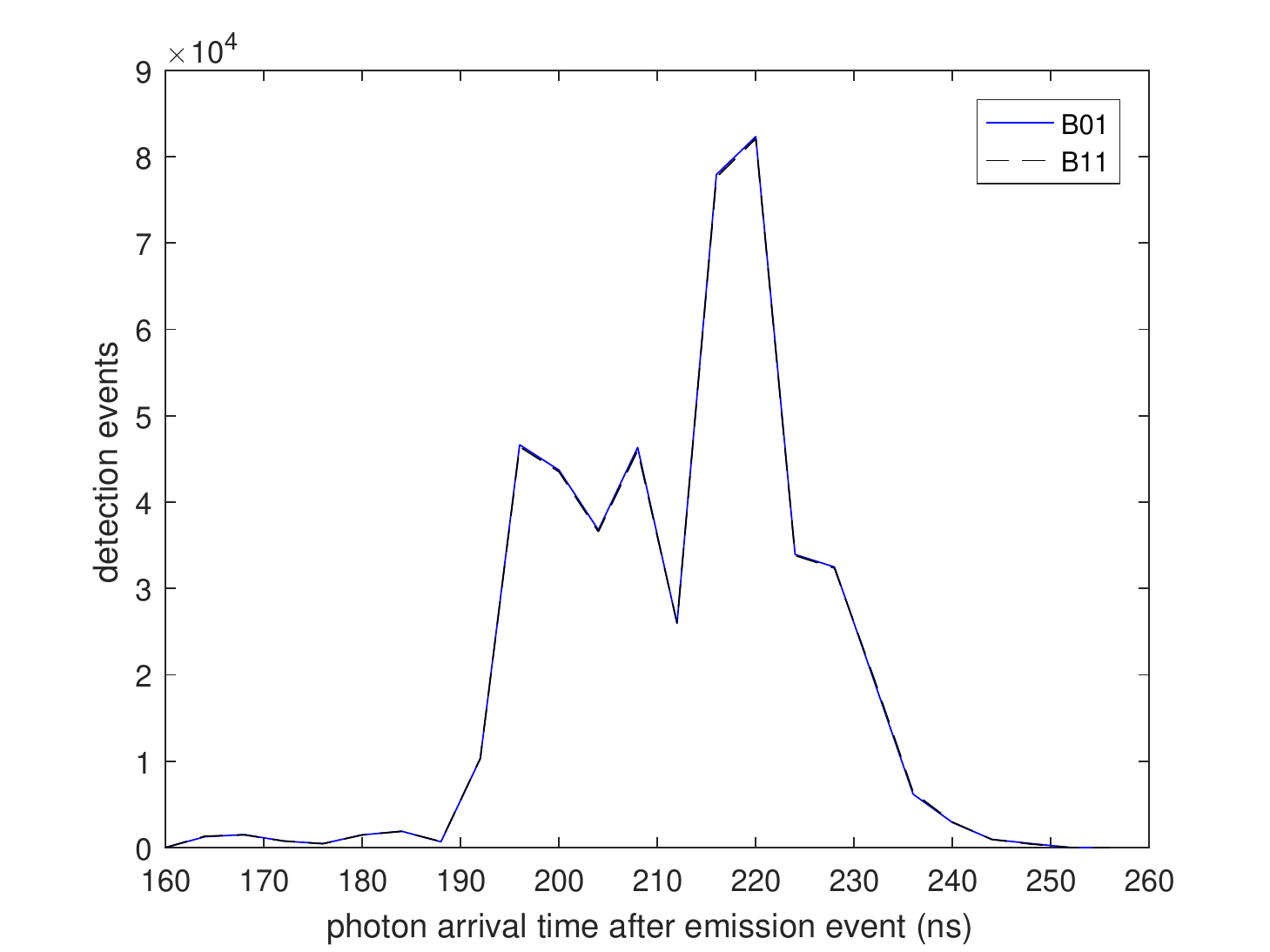}
\caption{Comparison of time-tags distribution of detections events depending on the other party choice in the Vienna experiment. Notation: $Xab$ means detection at party $X=A,B$ and corresponding choices $a,b=0,1$ by observers $A,B$, respectively. Each bin covers $4$ns counted with respect to the start signal. Note that both plots are almost overlapping.}\label{hist-v}
\end{figure}

 Due to smaller setup the statistics is larger and therefore also violation of local realism results in lower $P$-value. As in NIST the filtered data have been tested for signaling \cite{vien} and the table below shows actual event numbers in Table \ref{tabv} (choices $01$ correspond to $12$ in \cite{vien}) and also with invalid outcomes counted as detection event ($=1$) in Table \ref{tabv2}.
The $P$-value for no-signaling is everywhere rather large, $\sim 1$. Additionally, one can examine exact times of detection of photons, tagged in $250$ bins of $4$ns  \cite{mgius}. As shown in Fig. \ref{hist-v}, no significant difference appears between different choices of the other party (i.e. detection at $A$ does not depend on the choice of $B$) in comparison to the Poisson standard deviation $\sqrt{N_{bin}}\sim 300$ ($N_{bin}$ number of registered clicks $=1$ in a particular bin).
Hence, the distribution of detection times does not show violation of no-signaling within the available data.

However, in  a simple model of the measurement the time distributions for different choices should be identical up to a scaling factor.
Namely, a simple entangled state reads
\begin{equation}
|\psi\rangle=\alpha|+_A-_B\rangle+\beta|-_A+_B\rangle\label{een}
\end{equation}
with complex coefficients $\alpha$ and $\beta$ such that $|\alpha|^2+|\beta|^2=1$.
in the space of respective observers $|\pm_A\pm_B\rangle$. If the observables are constructed from  local pseudospin operators
$O=|+\rangle\langle +|-|-\rangle\langle -|$ and free chosen local unitary transformations $U_{Xn}|\psi\rangle$ (for a choice $Xn$, $X=A,B$) then quantum mechanics predicts only single parameter,
probability of detection of $+$ or $-$. Then the distribution of detection times can be arbitrary for both observers (depending e.g. on the unknown details of the detection setup) but can change only by a \emph{single} scaling factor (the total probability) for all tags when changing the choice, i.e. $p(t,A=0)/p(t,A=1)$ is a constant for all time tag bins $t$.

\begin{figure}
\includegraphics[scale=.5]{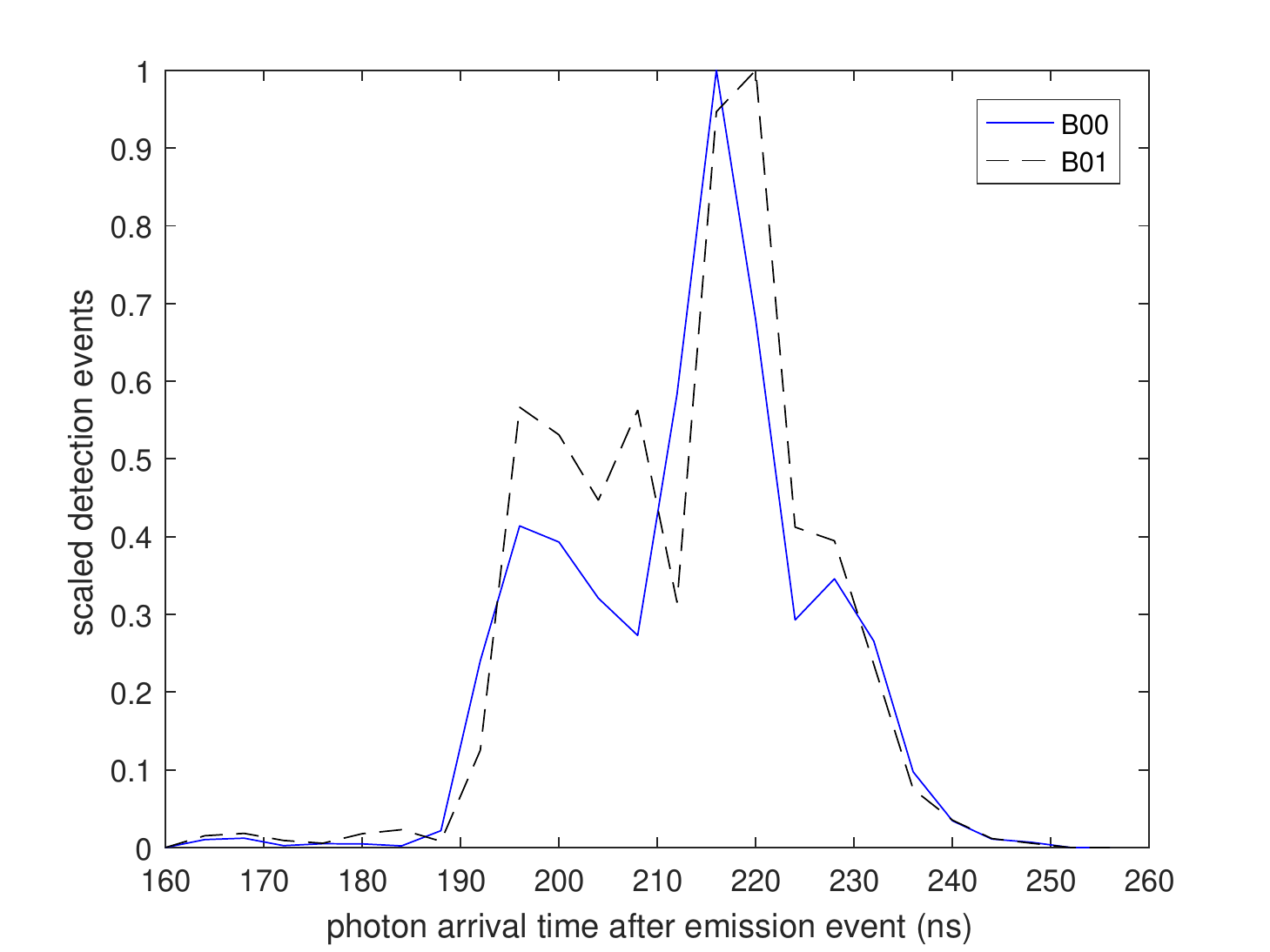}
\caption{Comparison of plots of $B00$ and $B01$ from Fig. \ref{hist-v} rescaled to their maxima.}\label{comp}
\end{figure}

However, as shown in Fig. \ref{comp} the distributions rescaled to their maxima are not the same. Quantitatively
after the rescaling the ration $B00/B01$ is estimated as $87\%$ while it should be $100\%$ (with Poisson error less than $ 1\%$)
Even if the detection times are sensitive to polarization, the reported visibility above $99\%$ excluded the residual contribution from
false polarization.

\begin{table}

\begin{tabular}{c|c|c|c}
$A=1$&$0$&$1$&$a$\\
\hline
$0$&$214830$&$583405 $&\\
\cline{1-3}
$1$&$214772$&$584837$&\\
\cline{1-3}
$b$&\multicolumn{3}{c}{}\\

\end{tabular}
\begin{tabular}{c|c|c|c}
$B=1$&$0$&$1$&$a$\\
\hline
$0$&$217663$&$217080$&\\
\cline{1-3}
$1$&$473599$&$472377$&\\
\cline{1-3}
$b$&\multicolumn{3}{c}{}\\
\end{tabular}
\caption{Number of detection events $(A,B=1)$ for all combinations of $a$ and $b$ times for the available data of Vienna experiment.}\label{tabv}
\end{table}

\begin{table}

\begin{tabular}{c|c|c|c}
$A=1$&$0$&$1$&$a$\\
\hline
$0$&$215005$&$583870 $&\\
\cline{1-3}
$1$&$214929$&$585292$&\\
\cline{1-3}
$b$&\multicolumn{3}{c}{}\\

\end{tabular}
\begin{tabular}{c|c|c|c}
$B=1$&$0$&$1$&$a$\\
\hline
$0$&$217949$&$217345$&\\
\cline{1-3}
$1$&$474160$&$472960$&\\
\cline{1-3}
$b$&\multicolumn{3}{c}{}\\
\end{tabular}
\caption{Number of detection events $(A,B=1)$ for all combinations of $a$ and $b$ times for the available data of Vienna experiment with invalid outcome included.}\label{tabv2}
\end{table}

Of course, the linear independence of the distributions can be explained within quantum mechanics by expanding the space of states. For instance, modifying the initial state $|\psi\rangle\to |\psi\rangle|0_A0_B\rangle$
and the choices $U_n\to U_n (|n\rangle\langle 0|+|0\rangle\langle n|)$. Then we have more final states, i.e. $|\pm n\rangle$ and the state $n$ may lead to a different distribution of tags. Technically, the choice is realized by Pockels cells which turn the relative phase of the photon (in the $\pm$ space) depending on the driving electric field (whose strength is freely chosen by the bit from random number generator). If the Pockels cell creates an additional state (e.g. by emitting an auxiliary photon or other particle or considerably changing the path of the photon) then the photodetection time may start to depend on it. Even if this indeed happens, it means that the process of quantum photodetection is not yet fully understood and the further such experiments may help to describe the mechanics of Pockels cells. Estimating Pockels cell length at $2$cm and the length of the photon at $3$m and high visibility, the effect must be quite exotic. The Bell example violating local realism is completely described within $2\times 2$ space. The physical space is of course larger but the other degrees of freedom should decouple. Note that the detection process ultimately maps the Bell state onto large macroscopic pointer basis but it is rather strange when the other degrees of freedom become relevant already during the choice. This effect should be identified by repeating the experiment in different regimes.

\begin{figure}
\includegraphics[scale=.4]{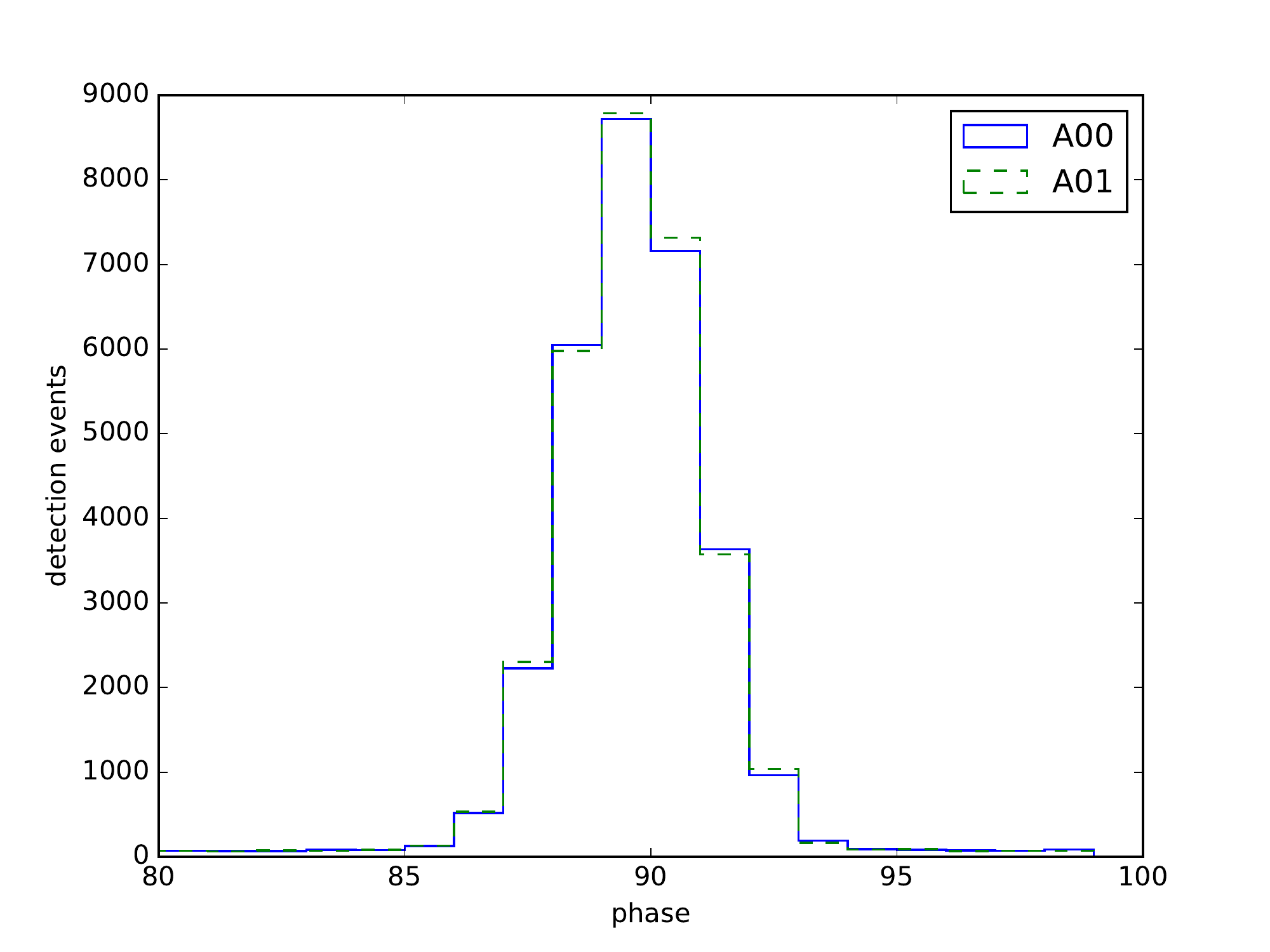}
\includegraphics[scale=.4]{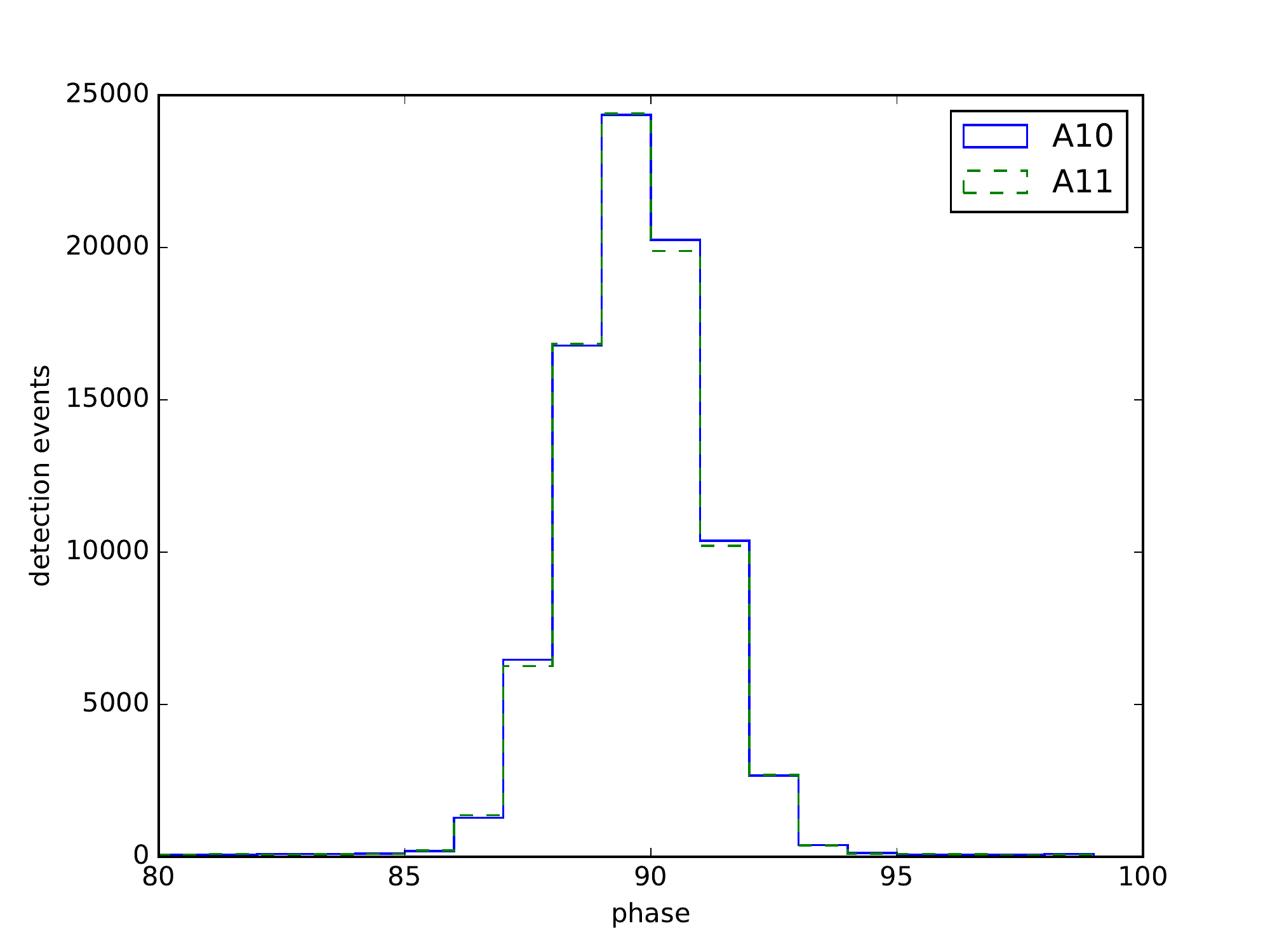}
\includegraphics[scale=.4]{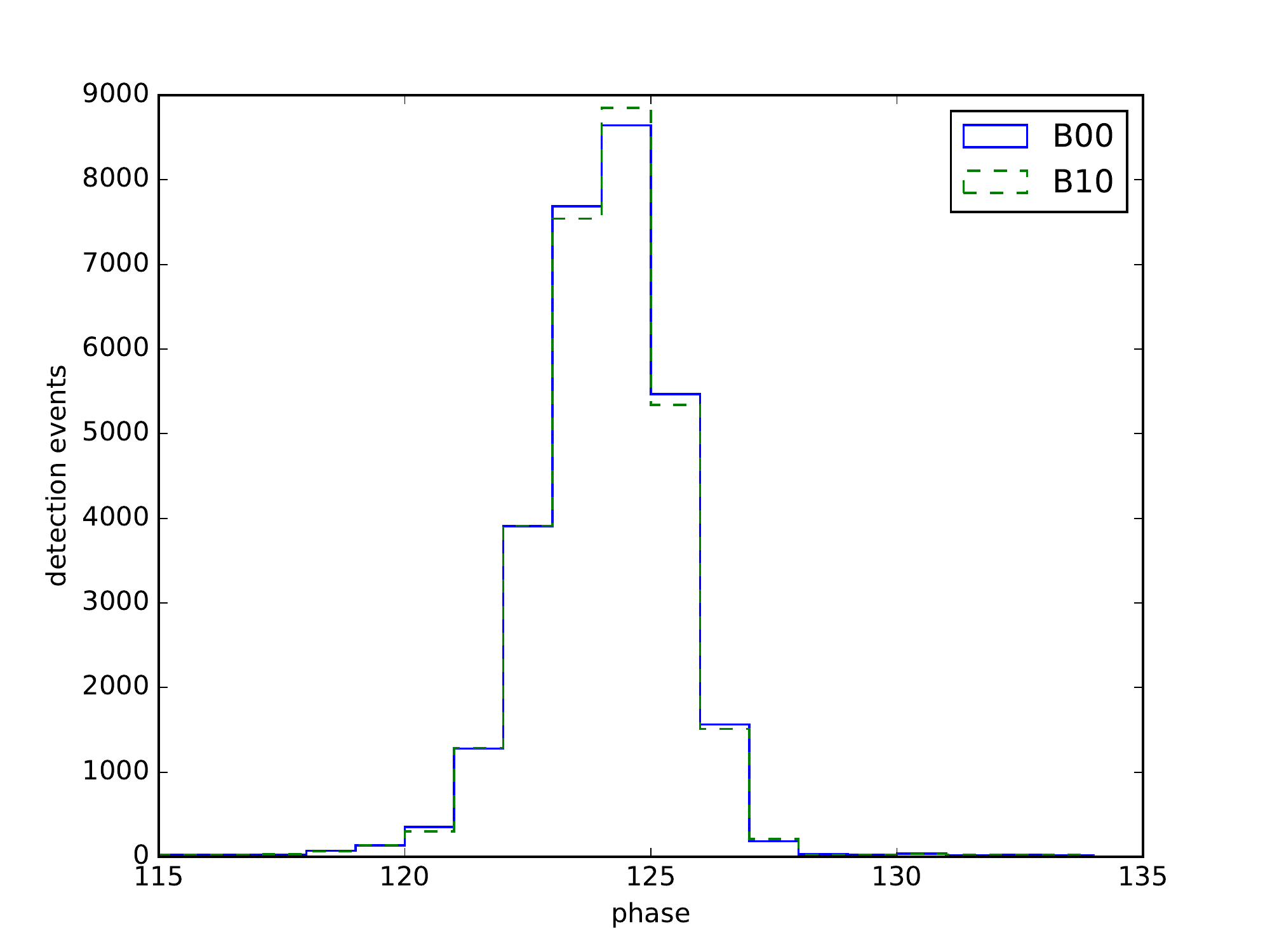}
\includegraphics[scale=.4]{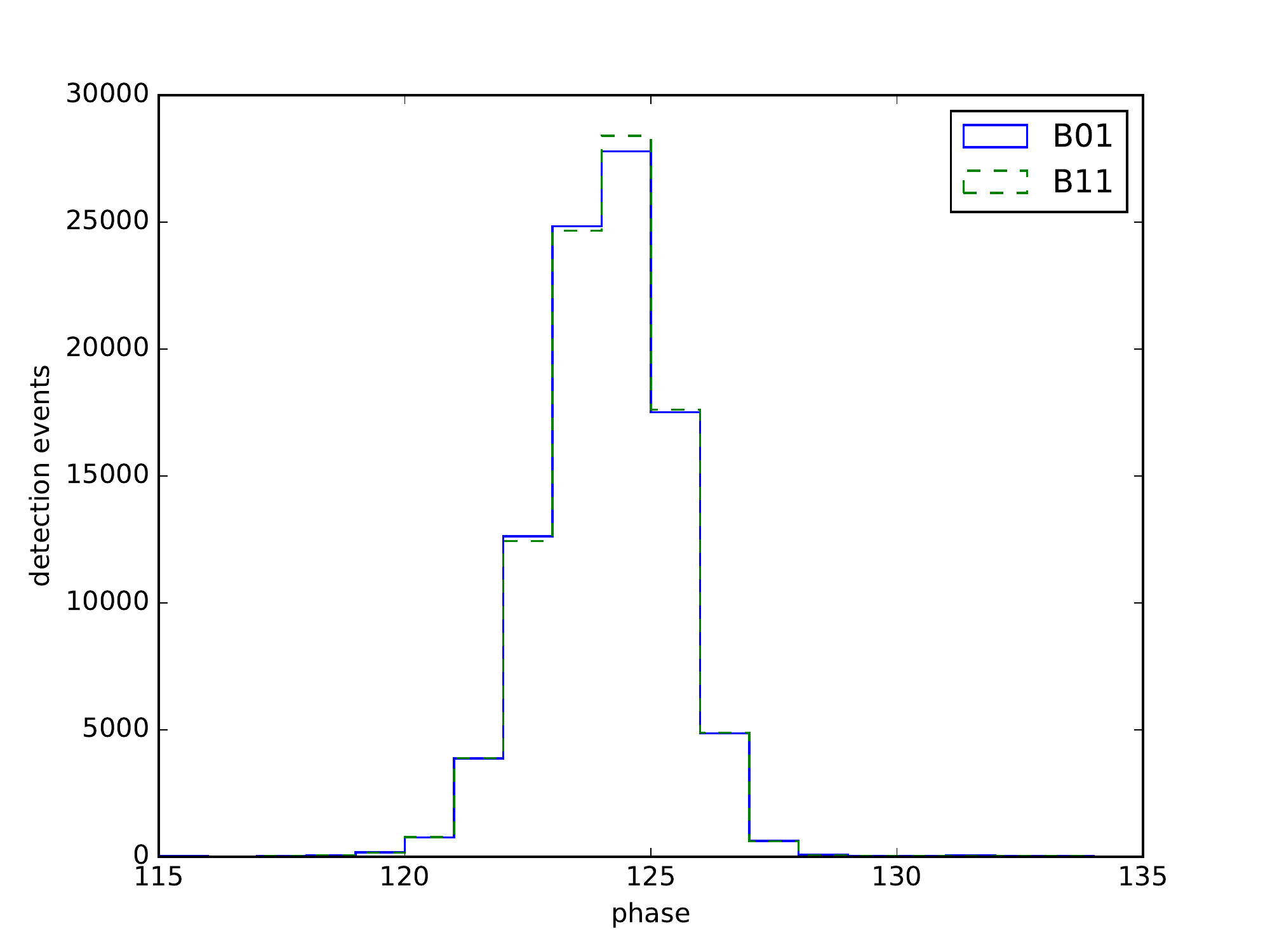}
\caption{Comparison of time-tags distribution of detections events depending on the other party choice in the NIST experiment for pulses $1-15$ counted from the start of the chosen rotation ($28$ for A and $37$ for B). Notation
as in Fig. \ref{hist-v}}\label{hnist}
\end{figure}

Note that such effect is not visible in the NIST experiment, taking into account only pulses affected by Pockels cells ($1-15$ counted from $28$ for A and $37$ for B), see Fig. \ref{hnist} (for run XOR3;  results are similar for runs XOR2 and 1).

Summarizing the Vienna experiment, it is in perfect agreement with no-signaling assumption. However, the data are inconsistent with the simplest $2\times 2$ quantum space, possibly due to additional states created in Pockels cells. This effect should help in finding a correct quantum mechanical description of the actual experiments. This discrepancy does not invalidate the violation of local realism, which is not restricted to a particular quantum model.

\section{Munich experiment}

The setup in Munich \cite{munch} is similar to the Delft one, again as in Fig. \ref{sii}.
Here the spatial/temporal distance $x-v$ and $y-u$ is 398m/1$\mu$s (spacelike). The time of $z$ ($C$ measurement) occurs before $u,v$ so before choices $a$ and $b$. The experiment also involves entanglement swapping.
First a two-level atom is entangled with a photon, then photons swap the entanglement to atoms. The successful swapping is heralded by $C=1$ is a coincidence of superposed photons clicks. 
The choices $a$ and $b$ imply different ionizing photons polarization, corresponding to  a basis in the two-level space. The readout is 
$A=+1$ or $-1$ refers to the detected ionized state or not ionized state, respectively.

The heralding ($C=1$) preselects one of the Bell states
\begin{equation}
\sqrt{2}|\psi_\pm\rangle=|+-\rangle\pm|-+\rangle
\end{equation}
Here $\pm$ depend on conditions required from the the coincidence. One can treat each case separately.
The correlations measurements are made in standard basis, namely $\langle \sigma_a\sigma_b\rangle$
with $\sigma_a=e^{ia}|+\rangle\langle -|+\mathrm{h.c.}$ and similarly $b$. The setting angles used in \cite{munch}
are $a=0$, $a'=\pi/2$, $b=-\pi/4$, $b'=\pi/4$. In all these cases ideal quantum theory predicts $\langle \sigma_a\sigma_b\rangle=\pm 1/\sqrt{2}$. Certainly finite efficiency and visibility can scale or shift this correlation but only by a common factor for all settings.
This is obviously violated, taking the data from \cite{munch} Fig. 3 and Table S7.
For instance for $\psi_+$ one reads 
$\langle \sigma_{a'}\sigma_b\rangle=-0.603\pm0.022$ while $\langle\sigma_{a'}\sigma_{b'}\rangle=0.463\pm 0.025$ giving the difference $0.14$ beyond $4$ standard deviations. Of course, the explanation -- as in Vienna -- can be that  the setting
affects the efficiency or even change the state into some non-maximally entangled one. This is technically plausible but goes beyond the simplest Bell model of state and measurement.

One should continue or repeat such experiment to identify all important effects outside of the Bell model, e.g. check if the preselected state is indeed non-maximally entangled and why. One can perform the similar analysis as in Delft \cite{hensen}
where the state has been found  non-maximally entangled which led to optimization of choice angles $a,a',b,b'$. Let us also stress that this effect has nothing to do with violation of no-signaling, which is not observed in Munich, and does not invalidate violation of local realism.

\section{Freedom and randomness of choice}

All the four experiments rely at least partly on quantum random number generators which are similar to each other \cite{morgan}.
The problem of such random choice is that the bits generated in a quantum process must become classical before affecting the entangled state.
In Delft and Munich the distance is so large that one can rather trust classicality of the random numbers. In NIST the distance is smaller but the quantum outcome is combined with a pseudorandom number taken artificially from a cultural source (e.g. a movie).
In Vienna, due to the short distance one has to trust that the quantum generator indeed immediately makes the outcomes classical.
However, in all experiments from a superdeterminist point of view one should treat the quantum generator as a part of the system.
Suppose the generator produces states $|G\rangle=(|0\rangle+|1\rangle)/\sqrt{2}$ and one measures whether it is in the state $0$ or $1$.
It will certainly pass all sophisticated tests for randomness but still will change the perspective on local realism. Now the complete initial entangled state (\ref{een}) should rather read $|G_AG_B\psi\rangle$ and the local unitary transformations read
$U_0|0\rangle\langle 0|+U_1|1\rangle\langle 1|$. Now one can measure \emph{jointly} if the system is in a state $n\pm$, $n=0,1$ which gives the statistics apparently violating local realism, although there was no choice at all. 

The random choice is electronic and in principle could have been predetermined in the past common for all observes.
The common past could be only partly avoided by relying on cosmic radiation \cite{cosm}(although with detection loophole) or
relying on human operation \cite{wise}. However, in principle even human choice could be predetermined and we have to take a pragmatic point of view (rejecting superdeterminism)-- trust the random choice hypothesis at a certain level.
Such trust is an important assumption of all the four experiments and can be quantified by comparing margins between 
the reach of the relativistic signal from the random number generator of the other party and the completion of the readout. We have summarized these margins in Table \ref{tabsum}. Increasing the margin in the future may help to test, e.g., different methods of random numbers generation like collecting them from cultural artifacts \cite{piron}, with additional randomness security such as cloning the numbers and storing them independently of the main experiment, preventing their uncontrolled leakage during the run.
The readout should also be cloned or copied to be secured from hacking before entering the other party's choice lightcone.
The $P$-value for local realism should in principle be independent of the random number generation and data storage method for identical other conditions of the experiment.

\begin{table}
\begin{tabular}{c|c|c|c|c}
&Delft&NIST&Vienna&Munich\\
\hline
signalling&moderate&moderate&insignificant&insignificant\\
\cline{1-5}
simple model&sufficient&sufficient&insufficient&insufficient\\
\cline{1-5}
margin[ns]&90&25& 7& 314\\
\end{tabular}

\caption{Summary of the analysis of the experiments with respect to signaling, simple quantum model and margin for choice and readout. In NIST the the margin refers to the latest pulse considered in violation of local realism \cite{nist}.}\label{tabsum}
\end{table}

\section{Discussion}

As the authors of experiments \cite{hensen,nist,vien,munch} stress, all tests of local realism rely on assumptions that must be trusted
by faithful description of the setups. Otherwise one could invent endless loopholes for local realism. However, this trust can be also 
verified by careful analysis of the data. In the hereby analysis of these assumptions anomalies of moderate significance have been found
summarized in Table \ref{tabsum}.
The $P$-value for the particular no-signaling and independence test is similar to local realism in Delft and similar to the analysis of no-signaling in NIST with larger detection windows ($P\sim 5\%$). In NIST it is certainly much larger than the one for local realism. The Vienna and Munich experiments do not show violation of no signaling but cannot be described by the simplest quantum model. The present data are insufficient to refute no-signaling principle but further experiments and analysis may change the confidence level. Violation of no-signaling will be confirmed even if only one (but reproducible) experiment shows it with $P\sim 10^{-7}$ or less (rule of $5$ standard deviations). One should remember that the present quantum theoretical model does not predict signaling at all, therefore one should expect much smaller deviations than for violation of local realism. This is why the Bell-type experiments should be continued to check if there is even a tiny trace of signaling.
As regards relativity, the signaling may be only apparently superluminal because there could be effects changing  the factual time of the choice or readout (e.g. synchronization error). In NIST the examined time window covers also region reachable by signaling slower than light. Both choices and readouts are so far concluded by trust in fast electronics. Their time margin and random number generation methods should be improved in future \cite{wise,piron}.
The further analysis of the deviation from a simplified quantum model in Vienna and Munich should help in deriving more accurate model of optical quantum systems, although the deviation is irrelevant for the violation of local realism. 

\section*{Acknowledgements}
I thank L.K. Shalm, S. Glancy, P.L. Bierhorst, S.W. Nam, R. Hanson, M. Giustina, M. W. Mitchell, R. Demkowicz-Dobrza\'nski  and J.-A. Larsson for providing the data, helping in their processing and discussions.

\end{document}